\documentclass[pre,aps,preprint,showpacs,preprintnumbers,amsmath,amssymb,secnumroman,eqsecnum]{revtex4}

\usepackage{graphicx}% Include figure files
\usepackage{dcolumn}% Align table columns on decimal point
\usepackage{bm}% bold math

\newcommand{\beq}{\begin{equation}}
\newcommand{\eeq}{\end{equation}}
\newcommand{\beqa}{\begin{eqnarray}}
\newcommand{\eeqa}{\end{eqnarray}}
\newcommand{\vc}[1]{\mbox{\boldmath $#1$}}

\newcommand{\vol}[1]{{\bf #1}}

\begin{document}

%\preprint{APS/123-QED}

\title{Jittery velocity relaxation of an elastic sphere immersed in a viscous incompressible fluid}% Force line breaks with \\

\author{B. U. Felderhof}

 \email{ufelder@physik.rwth-aachen.de}
\affiliation{Institut f\"ur Theoretische Physik A\\ RWTH Aachen\\
Templergraben 55\\52056 Aachen\\ Germany\\
}%

\date{\today}% It is always \today, today,

\begin{abstract}
Velocity relaxation of an elastic sphere immersed in a viscous incompressible fluid is studied on the basis of the equations of linear elasticity and the linearized Navier-Stokes equations. It is found that both translational motion after a sudden impulse and rotational motion after a sudden twist show jittery behavior in the long-time regime, with many reversals of velocity if the sphere is sufficiently flexible. In the extreme long-time regime the translational and rotational velocity relaxation functions each decay with a universal algebraic long-time tail. The added mass and the added moment of inertia of the sphere are found to vanish.
\end{abstract}

\maketitle
\section{\label{I}Introduction}

The dynamics of an elastic sphere immersed in a viscous incompressible fluid is complicated due to the coupling of the vibrational sound modes of the sphere to the dissipative motion of the surrounding fluid. In the following we investigate in particular how the translational and rotational velocity of the sphere relax to zero after a sudden impulse or twist. For simplicity we assume a no-slip condition in the coupling of solid to fluid, but the calculations can be readily extended to the case of perfect slip, or to a mixed slip-stick boundary condition. For a rigid sphere with no-slip boundary condition the decay of velocity is monotonic, but the velocity of a sufficiently flexible sphere oscillates many times about zero, before finally decaying with a positive and universal algebraic long-time tail.

The frequency-dependence of both the translational and the rotational friction coefficient shows many resonances corresponding to the sound modes of the isolated sphere. The analytic form of the translational friction coefficient is so complicated that we do not give the explicit expression, but we indicate how it can be derived from a set of four linear equations for four coefficients characterizing the motion of fluid and solid.

We find that for an elastic sphere the added mass vanishes, as in the case of a permeable sphere \cite{1}. In general, the added mass is defined from the high frequency behavior of the friction coefficient, as the coefficient of the behavior linear in frequency. For a rigid sphere the added mass equals one half of the displaced fluid mass \cite{2},\cite{3}. For an elastic sphere the friction coefficient tends to a constant as the frequency tends to infinity in the upper half of the complex plane. The constant is independent of viscosity and mass density of the fluid, but does depend on elastic constants, mass density, and radius of the sphere. At low frequency the friction coefficient depends only on fluid viscosity and sphere radius. In particular, the steady-state friction coefficient takes the Stokes value.

The frequency-dependent rotational friction coefficient is given by a relatively simple expression. The added moment of inertia vanishes, as in the case of a permeable sphere \cite{1}. Again the friction coefficient tends to a constant as the frequency tends to infinity in the upper half of the complex plane. The constant is independent of the properties of the fluid, but does depend on the properties of the sphere. The steady-state friction coefficient takes the value derived by Kirchhoff for a rigid sphere. This depends only on fluid viscosity and sphere radius.

The velocity relaxation function follows from the equation of motion for the sphere via the hydrodynamic force and torque exerted by the fluid. The dependence on time can be calculated by numerical Fourier transform of the complex admittance as a function of frequency. Due to the coupling to many sound modes the motion is complicated. The sound modes of the isolated sphere consist of standing longitudinal and transverse sound waves in a sphere with free surface boundary conditions, as well as Rayleigh surface modes \cite{4}. The jittery motion of the immersed sphere in the long-time regime can be calculated and observed, but is not easily characterized.

The relevance of elasticity to the overall motion of the sphere can be measured in terms of a dimensionless parameter expressing the ratio of characteristic time scales of fluid and solid. The parameter is proportional to the sphere radius, which indicates that elasticity is particularly important in micro-hydrodynamics. It is known that continuum hydrodynamics finds application on a surprisingly small length scale \cite{5}. It will be of interest to study the motion of small elastic spherical particles in experiment or computer simulation. On account of the fluctuation-dissipation theorem \cite{6} the time-correlation function of Brownian motion is proportional to the velocity relaxation function, so that the behavior found here can also be seen in Brownian motion. The observation of Brownian motion can therefore be used as a tool to investigate the elastic properties of suspended particles.

Our study is of particular relevance to soft matter physics. Although Young's modulus is of the order of $10^{11}$ Pa for metals, and is more than $10^{12}$ Pa for diamond, it can be many orders of magnitude smaller in soft matter. One can compare the characteristic time $\tau_{VY}$, defined as the ratio of the shear viscosity of the liquid to Young's modulus of the solid, with a typical viscous relaxation time $\tau_v$. For a soft solid the time $\tau_{VY}$ can become of the order of the viscous relaxation time $\tau_v$, and then elasticity must be taken into account. For equal mass density of liquid and solid the ratio of $\tau_{VY}$ to viscous relaxation time $\tau_v$ is approximately the square of the dimensionless parameter mentioned above.

For simplicity we take the fluid to be incompressible. Since in most practical applications the fluid will be water, this should be an excellent approximation. In principle, fluid compressibility can be taken into account \cite{7}, but at the expense of more parameters and a more elaborate calculation.

\section{\label{II}Translational motion}

We consider an elastic sphere of radius $a$, immersed in a viscous incompressible fluid of shear viscosity $\eta$ and mass density $\rho$. The flow velocity of the fluid $\vc{v}(\vc{r},t)$ and the pressure $p(\vc{r},t)$ are assumed to satisfy the linearized Navier-Stokes equations \cite{8}
\begin{equation}
\label{2.1}\rho\frac{\partial\vc{v}}{\partial
t}=\eta\nabla^2\vc{v}-\nabla
p,\qquad\nabla\cdot\vc{v}=0,
\end{equation}
for $r>a$. The flow velocity $\vc{v}$ is assumed to vanish at infinity and the pressure $p$ tends to the constant ambient pressure $p_0$. The elastic displacement $\vc{u}(\vc{r},t)$ of the sphere is assumed to satisfy the linear equation of motion\cite{4}
\begin{equation}
\label{2.2}\rho_s\frac{\partial^2\vc{u}}{\partial
t^2}=\mu\nabla^2\vc{u}+(\frac{1}{3}\mu+\kappa)\nabla\nabla\cdot\vc{u}+\vc{F}(t),
\end{equation}
for $r<a$, where $\mu$ is the shear modulus, $\kappa$ is the compression modulus, and $\vc{F}(t)$ is a uniform time-dependent force density applied to the sphere. We assume that the system is at rest for $t<0$ and is set in motion by an impulsive force acting at $t=0$, so that
\begin{equation}
\label{2.3}\vc{F}(t)=\frac{3}{4\pi a^3}\vc{P}\delta(t),
\end{equation}
where $\vc{P}$ is the imparted impulse.

The solution of Eq. (2.1) is found most easily by a Fourier transform with respect to time. Thus we define Fourier components
$\vc{v}_\omega(\vc{r})$ and $p_\omega(\vc{r})$ by
\begin{equation}
\label{2.4}\vc{v}_\omega(\vc{r})=\int^\infty_0e^{i\omega
t}\vc{v}(\vc{r},t)\;dt,\qquad p_\omega(\vc{r})=\int^\infty_0
e^{i\omega t}p(\vc{r},t)\;dt.
\end{equation}
The equations for the Fourier components read
\begin{equation}
\label{2.5}\eta[\nabla^2\vc{v}_\omega-\alpha^2\vc{v}_\omega]-\nabla p_\omega=0,\qquad\nabla\cdot\vc{v}_\omega=0,\qquad\mathrm{for}\;r>a,
\end{equation}
with complex wavenumber
\begin{equation}
\label{2.6}\alpha=(-i\omega\rho/\eta)^{1/2},\qquad
\mbox{Re}\;\alpha>0.
\end{equation}
Similarly, inside the sphere the Fourier component
\begin{equation}
\label{2.7}\vc{u}_\omega(\vc{r})=\int^\infty_0e^{i\omega
t}\vc{u}(\vc{r},t)\;dt
\end{equation}
satisfies the equation
\begin{equation}
\label{2.8}\mu[\nabla^2\vc{u}_\omega+q_t^2\vc{u}_\omega]+(\frac{1}{3}\mu+\kappa)\nabla\nabla\cdot\vc{u}_\omega=-\hat{\vc{F}},\qquad\mathrm{for}\;r<a,
\end{equation}
where
\begin{equation}
\label{2.9}q_t=\frac{\omega}{c_t},\qquad c_t=\sqrt{\frac{\mu}{\rho_s}},\qquad\hat{\vc{F}}=\frac{3}{4\pi a^3}\vc{P}.
\end{equation}
Here $q_t$ is the wavenumber of transverse sound waves at frequency $\omega$.

The equations must be solved under the conditions that $\vc{v}_\omega$ and $p_\omega$ tend to zero at infinity, and that the fluid velocity just outside the sphere equals the time-derivative of the displacement at the sphere surface. In addition the normal-normal and normal-tangential components of the stress tensor must be continuous at $r=a$. The stress tensor of the fluid is given by
\begin{equation}
\label{2.10}\vc{\sigma}_f=\eta(\nabla\vc{v}+\widetilde{\nabla\vc{v}})-p\vc{I},
\end{equation}
where $\vc{I}$ is the unit tensor. The stress tensor of the solid is given by
\begin{equation}
\label{2.11}\vc{\sigma}_s=\mu(\nabla\vc{u}+\widetilde{\nabla\vc{u}}-\frac{2}{3}\vc{I}\nabla\cdot\vc{u})+\kappa\vc{I}\nabla\cdot\vc{u}.
\end{equation}

We choose coordinates such that the $z$ axis is in the direction of the impulsive applied force. Then, with spherical coordinates $(r,\theta,\varphi)$, it suffices to consider the radial and polar components of the fluid velocity $v_r,v_\theta$ and of the displacement $u_r,u_\theta$, as well as the $rr$ and $r\theta$ components of the stress tensor at the interface of fluid and solid.

The resulting center of mass velocity of the sphere $\vc{U}_\omega$ is given by
\begin{equation}
\label{2.12}\vc{U}_\omega=\frac{-3i\omega}{4\pi a^3}\int_{r<a}\vc{u}_\omega\;d\vc{r}.
\end{equation}
By symmetry this is also in the $z$ direction. The fluid equations of motion can be reduced to scalar form by the Ansatz
\begin{eqnarray}
\label{2.13}\vc{v}_\omega(\vc{r})&=&f_A(r)\vc{e}_z+f_B(r)(\vc{I}-3\hat{\vc{r}}\hat{\vc{r}})\cdot\vc{e}_z,\nonumber\\
p_\omega(\vc{r})&=&\eta g(r)\hat{\vc{r}}\cdot\vc{e}_z,
\end{eqnarray}
where the subscripts $A$ and $B$ refer to the two types of vector spherical harmonics that come into play \cite{2}. We find that the radial functions must satisfy the coupled equations
\begin{eqnarray}
\label{2.14}\frac{d^2f_A}{dr^2}+\frac{2}{r}\frac{df_A}{dr}-\alpha^2f_A-\frac{1}{3}\bigg(\frac{dg}{dr}+\frac{2}{r}\;g\bigg)=0,\nonumber\\
\frac{d^2f_B}{dr^2}+\frac{2}{r}\frac{df_B}{dr}-\frac{6}{r^2}f_B-\alpha^2f_B+\frac{1}{3}\bigg(\frac{dg}{dr}-\frac{1}{r}\;g\bigg)=0,\nonumber\\
\frac{d^2g}{dr^2}+\frac{2}{r}\frac{dg}{dr}-\frac{2}{r^2}\;g=0,\qquad\mathrm{for}\;r>a.
\end{eqnarray}
The radial and polar components of the flow velocity are
\begin{equation}
\label{2.15}v_r(r,\theta)=(f_A-2f_B)\cos\theta,\qquad v_\theta(r,\theta)=-(f_A+f_B)\sin\theta,
\end{equation}
and $v_\varphi=0$. The relevant non-vanishing components of the fluid stress tensor are
\begin{eqnarray}
\label{2.16}\sigma_{frr}(r,\theta)&=&\eta\big(2f'_A-4f'_B-g\big)\cos\theta,\nonumber\\
\sigma_{fr\theta}(r,\theta)&=&\sigma_{\theta r}(r,\theta)=\eta\big(-f'_A-f'_B+\frac{3}{r}f_B\big)\sin\theta,
\end{eqnarray}
where the prime denotes the derivative with respect to $r$. The solution of Eq. (2.14) takes the form \cite{9}
\begin{eqnarray}
\label{2.17}f_A(r)&=&2B_Nk_0(\alpha r),\qquad f_B(r)=B_Nk_2(\alpha r)-B_P/r^3,\nonumber\\
g(r)&=&B_P\frac{\alpha^2}{r^2},\qquad\mathrm{for}\;r>a,
\end{eqnarray}
with coefficients $B_N,B_P$ and modified spherical Bessel functions \cite{10} $k_l(z)$.

The radial and polar components of the displacement in the solid take the explicit form
\begin{eqnarray}
\label{2.18}u_r(r,\theta)&=&\big[A_N\big(2j_0(q_tr)+2j_2(q_tr)\big)+A_P\big(j_0(q_lr)-2j_2(q_lr)\big)+\hat{F}\big]\cos\theta,\nonumber\\
u_\theta(r,\theta)&=&\big[A_N\big(-2j_0(q_tr)+j_2(q_tr)\big)+A_P\big(-j_0(q_lr)-j_2(q_lr)\big)-\hat{F}\big]\sin\theta,
\end{eqnarray}
with coefficients $A_N,A_P$ and spherical Bessel functions \cite{10} $j_l(z)$. Furthermore $q_l$ is the wavenumber of longitudinal sound waves at frequency $\omega$ with sound velocity $c_l$,
\begin{equation}
\label{2.19}q_l=\frac{\omega}{c_l},\qquad c_l=\sqrt{\frac{3\kappa+4\mu}{3\rho_s}}.
\end{equation}

The boundary conditions can be applied at the undisplaced spherical surface. The conditions read
\begin{equation}
\label{2.20}v_r(a+,\theta)=-i\omega u_r(a-,\theta),\qquad v_\theta(a+,\theta)=-i\omega u_\theta(a-,\theta),
\end{equation}
together with
\begin{equation}
\label{2.21}\sigma_{frr}(a+,\theta)=\sigma_{srr}(a-,\theta),\qquad\sigma_{fr\theta}(a+,\theta)=\sigma_{sr\theta}(a-,\theta)
\end{equation}
yield four equations for the four coefficients $A_N,A_P,B_N,B_P$. These can be solved in algebraic form, but the resulting expressions are complicated.

\section{\label{III}Sphere velocity}

From the flow pattern calculated above we can evaluate the hydrodynamic force exerted by the fluid on the sphere.
From the linearized Navier-Stokes equations Eq. (2.1) it follows that the force can be expressed as the integral of the stress tensor over a spherical surface just outside the sphere,
 \begin{equation}
\label{3.1}\vc{K}_\omega=\int_{S(a+)}\vc{\sigma}_{f\omega}\cdot\hat{\vc{r}}\;dS.
\end{equation}
From Eqs. (2.16) and (2.17) we find $\vc{K}_\omega=K_\omega\vc{e}_z$ with
 \begin{equation}
\label{3.2}K_\omega=-\frac{4\pi}{3}\eta\big[6\alpha a^2k_1B_N+\alpha^2B_P\big].
\end{equation}
From Eq. (2.12) we find for the sphere velocity $\vc{U}_\omega=U_\omega\vc{e}_z$ with
 \begin{equation}
\label{3.3}U_\omega=-i\omega\bigg[6\frac{j_1(q_ta)}{q_ta}A_N+3\frac{j_1(q_la)}{q_la}A_P+\hat{F}\bigg].
\end{equation}
The translational friction coefficient $\zeta_T(\omega)$ is defined by
 \begin{equation}
\label{3.4}K_\omega=-\zeta_T(\omega)U_\omega.
\end{equation}

The sphere velocity is determined by the equation of motion
\begin{equation}
\label{3.5}-i\omega m_p\vc{U}_\omega=\vc{K}_\omega+\vc{E}_\omega,
\end{equation}
where $m_p=4\pi\rho_sa^3/3$ is the mass of the sphere and $\vc{E}_\omega$ is the applied external force. In the present case
$\vc{E}_\omega=\vc{P}$, independent of frequency $\omega$.  From Eq. (3.5) we find
\begin{equation}
\label{3.6}\vc{U}_\omega=\mathcal{Y}_T(\omega)\vc{E}_\omega,
\end{equation}
where $\mathcal{Y}_T(\omega)$ is the translational admittance, given by
\begin{equation}
\label{3.7}\mathcal{Y}_T(\omega)=[-i\omega m_p+\zeta_T(\omega)]^{-1}.
\end{equation}

We define the translational velocity relaxation function $\gamma_T(t)$ by
\begin{equation}
\label{3.8}U(t)=\frac{P}{m^*}\;\gamma_T(t),\qquad t>0,
\end{equation}
with effective mass $m^*=m_p+m_a$, where $m_a$ is the added mass. The relaxation function is related to the admittance by
\begin{equation}
\label{3.9}\int^\infty_0e^{i\omega t}\gamma_T(t)\;dt=m^*\mathcal{Y}_T(\omega).
\end{equation}
Hence the relaxation function $\gamma_T(t)$ can be evaluated by inverse Fourier transform.
It has the properties
\begin{equation}
\label{3.10}\gamma_T(0+)=1,\qquad\int^\infty_0\gamma_T(t)\;dt=\frac{m^*}{\zeta_T(0)}.
\end{equation}
In general the added mass of the particle is defined by
\begin{equation}
\label{3.11}m_a=\lim_{\omega\rightarrow\infty}\frac{\zeta_T(\omega)}{-i\omega}.
\end{equation}
In the present case the added mass vanishes.

The net displacement of the sphere depends only on the initial momentum and the steady-state friction coefficient,
\begin{equation}
\label{3.12}\vc{\Delta}_T=\int^\infty_0\vc{U}(t)\;dt=\vc{P}/\zeta_T(0).
\end{equation}
We find by series expansion of the friction coefficient in powers of $x=\alpha a$
\begin{equation}
\label{3.13}\zeta_T(\omega)=6\pi\eta a(1+x)+O(x^2).
\end{equation}
The first term shows that the steady-state friction coefficient has the Stokes value $\zeta_T(0)=6\pi\eta a$ independent of the elastic properties of the sphere.
The second term gives rise to the long-time behavior of the velocity,
\begin{equation}
\label{3.14}\vc{U}(t)\approx\frac{1}{12\rho(\pi\nu t)^{3/2}}\;\vc{P}\qquad\mathrm{as}\;t\rightarrow\infty,
\end{equation}
where $\nu=\eta/\rho$ is the kinematic viscosity. This is independent of the elastic properties and the mass of the sphere, and depends only on the properties of the fluid. At long times the particle moves with the fluid in uniform flow and is not deformed \cite{11}.

At high frequency, in the upper half complex plane, the friction coefficient tends to a constant given by
\begin{equation}
\label{3.15}\zeta_{T\infty}=\lim_{\omega\rightarrow\infty}\zeta_T(\omega)=\frac{4\pi}{3}(c_l+2c_t)\rho_sa^2.
\end{equation}
This depends only on the properties of the sphere.

\section{\label{IV}Analysis of velocity relaxation}

It is convenient to regard the friction coefficient as a function of the complex variable $x=\alpha a$, and to define the dimensionless admittance $\hat{F}(x)$ as
\begin{equation}
\label{4.1}\hat{F}_T(x)=\frac{4\pi m_p}{3m_f}\;\eta a\mathcal{Y}_T(\omega).
\end{equation}
Here we have chosen the pre-factor such that $\hat{F}_T(x)$ behaves as $1/x^2$ for large $x$. The function takes the form
\begin{equation}
\label{4.2}\hat{F}_T(x)=\frac{1}{x^2+MZ_T(x)},
\end{equation}
where we have abbreviated
\begin{equation}
\label{4.3}M=\frac{9m_f}{2m_p},\qquad\zeta_T(\omega)=6\pi\eta aZ_T(x).
\end{equation}
Here $m_f=4\pi\rho a^3/3$ is the mass of fluid displaced by the sphere.
In the present case the function $Z_T(x)$ is quite complicated. In comparison we have for a rigid sphere with no-slip boundary condition the simple quadratic expression \cite{2}
\begin{equation}
\label{4.4}Z_{Tns}(x)=1+x+\frac{1}{9}x^2.
\end{equation}
The quadratic behavior for large $x$ corresponds to the added mass $m_a=\frac{1}{2}m_f$.

Expanding the function $\hat{F}_T(x)$ in powers of $x$ we find
\begin{equation}
\label{4.5}Z_T(x)=1+x+Z_{T2}x^2+O(x^3),
\end{equation}
with coefficient
\begin{equation}
\label{4.6}Z_{T2}=\frac{1}{9}-\frac{\eta}{a\rho}\frac{c_l+c_t}{c_lc_t}+\frac{\eta^2}{a^2\rho}\frac{403\mu+210\kappa}{180\mu\kappa}.
\end{equation}
The derivation of this result is delicate. For example, it requires expansion of the numerator in the expression for the coefficient $B_N$ in powers of $x$ up to $x^{25}$. Note that in the limit $\kappa=b\mu\rightarrow\infty,\;\mu\rightarrow\infty$ with constant $b$, the coefficient tends to $1/9$, as in Eq. (4.4).

As in previous analysis \cite{12} we write the reduced admittance as a sum of simple poles in the complex $x$ plane of the form
\begin{equation}
\label{4.7}\hat{F}_T(x)=\sum_j\frac{A_j}{x-x_j}.
\end{equation}
Since the function behaves as $1/x^2$ at large $x$ one has the sum rules
\begin{equation}
\label{4.8}\sum_jA_j=0,\qquad\sum_jA_jx_j=1.
\end{equation}
In addition we find from Eq. (3.13)
\begin{equation}
\label{4.9}\sum_j\frac{A_j}{x_j}=\frac{-1}{M},\qquad\sum_j\frac{A_j}{x_j^2}=\frac{1}{M},
\end{equation}
Finally, Eq. (3.15) implies the sum rules
\begin{equation}
\label{4.10}\sum_jA_jx_j^2=0,\qquad\sum_jA_jx_j^3=-MZ_{\infty},\qquad Z_{\infty}=\frac{2}{9}(c_l+2c_t)\frac{\rho_s}{\eta}\;a,
\end{equation}
where $Z_{\infty}$ corresponds to $\zeta_{T\infty}$.

The relaxation function is given by
\begin{equation}
\label{4.11}\gamma_T(t)=\sum_jA_jx_jw(-ix_j\sqrt{t/\tau_v}),
\end{equation}
where $w(z)$ is the $w$ function \cite{10} $w(z)=\exp(-z^2)\mathrm{erfc}(-iz)$, and $\tau_v=a^2/\nu$ is the viscous relaxation time. The second sum rule in Eq. (4.8) corresponds to the initial value $\gamma_T(0+)=1$.

In our numerical work we consider a neutrally buoyant sphere with $m_f=m_p$, so that $M=9/2$. A convenient measure of the ratio of time scales of solid and fluid is given by the dimensionless parameter
\begin{equation}
\label{4.12}\sigma=(c_l+2c_t)\frac{\rho}{\eta}\;a.
\end{equation}
The limit $\sigma\rightarrow\infty$ corresponds to a rigid sphere. For small $\sigma$ the sphere is very floppy. With Young's modulus defined as $E=9\mu\kappa/(\mu+3\kappa)$ the characteristic time $\tau_{VY}$ is defined as $\tau_{VY}=\eta/E$. The ratio of times $\tau_v/\tau_{VY}$ is closely related to the parameter $\sigma$ with $\sigma\approx\sqrt{\rho\tau_v/\rho_s\tau_{VY}}$.

In the case of a rigid sphere with no-slip boundary condition the velocity relaxation function decays monotonically to zero. It can be expressed as an integral of exponentials $\exp(-ut/\tau_v)$ with a positive spectral density $p_T(u)$. The relaxation function is given by a sum of two terms of the form Eq. (4.11) with poles $x_{1ns},x_{1ns}^*$ with negative real part.

For the elastic sphere an infinite number of poles appears, corresponding to the sound modes of the sphere. The sound modes are oscillatory, but damped due to coupling to the viscous fluid. Correspondingly, some of the pole positions $\{x_j\}$ can be in the right-hand side of the complex $x$ plane, with imaginary part larger than the real part. The admittance is a complicated function of frequency, showing many resonances.

In Fig. 1 we show as an example $\mathrm{Re}\hat{F}_T(x)$ as a function of frequency for a sphere of radius $a=1$ with $\mu=\kappa=10\eta/\tau_v$, in a fluid of density $\rho=\rho_s=1$, corresponding to parameter $\sigma=11.16$, and compare with the corresponding function for a no-slip rigid sphere,
\begin{equation}
\label{4.13}\hat{F}_{Tns}(x)=\frac{1}{M+Mx+(m^*/m_p)x^2},
\end{equation}
with effective mass $m^*=m_p+\frac{1}{2}m_f$. For a neutrally buoyant sphere $M=9/2$ and $m^*/m_p=3/2$. For the chosen parameter values $Z_{T2}=-0.072$ and $Z_{\infty}=2.479$.

In Fig. 2 we plot the corresponding relaxation function $\gamma_T(t)$ as a function of $t/\tau_v$, as obtained by numerical inverse Fourier transform. We compare with the corresponding function $\gamma_{Tns}(t)$ for the no-slip sphere. This function has the initial value $\gamma_{Tns}(0+)=m_p/m^*=2/3$. For the above choice of parameters the relaxation function $\gamma_T(t)$ shows jittery motion with damped oscillations. We also compare with a sum of eleven terms of the form Eq. (4.11) corresponding to eleven poles and residues, with nine poles and residues calculated numerically from the exact expression for $\hat{F}_{T}(x)$, and the remaining two found by fitting the sum of pole terms to $\hat{F}_{T}(x)$ at $x=2.5,\;x=5,\;x=10$ and $x=20$. The last two conjugate poles and residues are necessary to reproduce the short-time behavior. Although the sum represents the relaxation function qualitatively reasonably well, it is clearly insufficient as an approximate description.

\section{\label{V}Rotational motion}

Next we consider rotational motion due to a sudden twist, corresponding to a time-dependent torque $\vc{N}(t)=\vc{L}\delta(t)$, applied to the sphere of moment of inertia $I_p=8\pi\rho_sa^5/15$, causing it to rotate and the fluid to move. The torque will be assumed small, so that we can again use linearized equations of motion. We shall be interested in calculating the time-dependent rotational velocity $\vc{\Omega}(t)$ of the sphere, as well as the corresponding flow pattern of the fluid.

We define Fourier components of the rotational velocity by
\begin{equation}
\label{5.1}\vc{\Omega}_\omega=\int^\infty_0e^{i\omega t}\vc{\Omega}(t)\;dt.
\end{equation}
The pressure remains constant and uniform, so that the equations for the Fourier components of the flow velocity read
\begin{equation}
\label{5.2}\eta[\nabla^2\vc{v}_\omega-\alpha^2\vc{v}_\omega]=0,\qquad\nabla\cdot\vc{v}_\omega=0,\qquad\mathrm{for}\;r>a.
\end{equation}
Inside the sphere
\begin{equation}
\label{5.3}\mu[\nabla^2\vc{u}_\omega+q_t^2\vc{u}_\omega]=-\hat{\vc{G}}\times\vc{r},\qquad\nabla\cdot\vc{u}_\omega=0,\qquad\mathrm{for}\;r<a,
\end{equation}
where
\begin{equation}
\label{5.4}\hat{\vc{G}}=\frac{15}{8\pi a^5}\vc{L}.
\end{equation}
The equations can be reduced to scalar form by the Ansatz
\begin{equation}
\label{5.5}\vc{v}_\omega(\vc{r})=f_C(r)\vc{e}_z\times\hat{\vc{r}},\qquad \vc{u}_\omega(\vc{r})=g_C(r)\vc{e}_z\times\hat{\vc{r}}.
\end{equation}
The radial functions $f_C(r)$ and $g_C(r)$ satisfy the equations
\begin{eqnarray}
\label{5.6}\mu\bigg[\frac{d^2g_C}{dr^2}+\frac{2}{r}\frac{dg_C}{dr}-\frac{2}{r}g_C+q_t^2g_C\bigg]=-\hat{G}r,\qquad\mathrm{for}\;r<a,\nonumber\\
\frac{d^2f_C}{dr^2}+\frac{2}{r}\frac{df_C}{dr}-\frac{2}{r}f_C-\alpha^2f_C=0,\qquad\mathrm{for}\;r>a.
\end{eqnarray}
In spherical coordinates the only non-vanishing component of the flow velocity and the displacement is
\begin{equation}
\label{5.7}v_\varphi(r)=f_C(r)\sin\theta,\qquad u_\varphi(r)=g_C(r)\sin\theta.
\end{equation}
The relevant components of the stress tensor are
\begin{eqnarray}
\label{5.8}\sigma_{fr\varphi}(r)=\sigma_{f\varphi r}(r)=\eta\bigg(f'_C-\frac{f_C}{r}\bigg)\sin\theta,\nonumber\\
\sigma_{sr\varphi}(r)=\sigma_{s\varphi r}(r)=\mu\bigg(g'_C-\frac{g_C}{r}\bigg)\sin\theta.
\end{eqnarray}
The solution of Eq. (5.6) takes the form
\begin{equation}
\label{5.9}g_C(r)=A_Mj_1(q_tr)-\frac{1}{\mu q_t^2}\;\hat{G}r,\qquad
f_C(r)=B_Mk_1(\alpha r).
\end{equation}
From the conditions that $v_\varphi=-i\omega u_\varphi $ at $r=a$, and that the $r\varphi$-components of the stress tensor are continuous, we find
\begin{eqnarray}
\label{5.10}A_M&=&\frac{\eta\omega}{\mu q_t^2}\;\frac{\alpha k_2(\alpha a)}{\eta\alpha\omega j_1(q_t a)k_2(\alpha a)-i\mu q_tj_2(q_t a)k_1(\alpha a)}\;\hat{G},\nonumber\\
B_M&=&\frac{\omega}{q_t}\;\frac{j_2(q_t a)}{\eta\alpha\omega j_1(q_t a)k_2(\alpha a)-i\mu q_tj_2(q_t a)k_1(\alpha a)}\;\hat{G}.
\end{eqnarray}

The hydrodynamic torque exerted by the fluid on the sphere is given by the integral
\begin{equation}
\label{5.11}\vc{T}_{\omega}=\int_{S(a+)}\vc{r}\times(\vc{\sigma}_{f\omega}\cdot\hat{\vc{r}})\;dS.
\end{equation}
This yields $\vc{T}_{\omega}=T_{\omega}\vc{e}_z$ with
 \begin{equation}
\label{5.12}T_{\omega}=-\frac{8\pi}{3}\eta a^3\alpha k_2(\alpha a)B_M.
\end{equation}
The rotational velocity of the sphere is defined as
 \begin{equation}
\label{5.13}\vc{\Omega}_{\omega}=-\frac{-15i\omega}{8\pi a^5}\int_{r<a}\vc{r}\times\vc{u}_\omega(\vc{r})\;d\vc{r}.
\end{equation}
Hence we find $\vc{\Omega}_{\omega}=\Omega_{\omega}\vc{e}_z$ with
 \begin{equation}
\label{5.14}\Omega_\omega=-i\omega\bigg[5\frac{j_2(q_ta)}{q_ta^2}A_M-\frac{\hat{G}}{\mu q_t^2a}\bigg].
\end{equation}
The rotational friction coefficient $\zeta_R(\omega)$ is defined by
 \begin{equation}
\label{5.15}T_{\omega}=-\zeta_R(\omega)\Omega_\omega.
\end{equation}
Correspondingly we define the dimensionless friction coefficient $Z_R(x)$ as
 \begin{equation}
\label{5.16}\zeta_R(\omega)=8\pi\eta a^3Z_R(x),
\end{equation}
and find by series expansion in powers of $x$
 \begin{equation}
\label{5.17}Z_R(x)=1+\bigg(\frac{1}{3}-\frac{3\eta^2}{7\rho\mu a^2}\bigg)x^2-\frac{1}{3}x^3+O(x^4).
\end{equation}
The first term shows that the steady-state friction coefficient takes the Kirchhoff value of a no-slip rigid sphere, $\zeta_R(0)=8\pi\eta a^3$. Corresponding to Eq. (5.17) we write the low-frequency expansion
of the friction coefficient as
\begin{equation}
\label{5.18}\zeta_R(\omega)=\zeta_R(0)+\zeta_{R2}\alpha^2a^2-\frac{1}{24\pi\eta}\;\zeta_R(0)^2\alpha^3+O(\alpha^4).
\end{equation}
At high frequency, in the upper half complex plane, the friction coefficient tends to a constant given by
\begin{equation}
\label{5.19}\zeta_{R\infty}=\frac{8\pi}{3}\;c_t\rho_sa^4,
\end{equation}
independent of viscosity.

The rotational velocity of the sphere is determined by the equation of motion
\begin{equation}
\label{5.20}-i\omega I_p\vc{\Omega}_\omega=\vc{T}_{\omega}+\vc{N}_\omega,
\end{equation}
where $\vc{N}_\omega$ is the external mechanical torque applied to the sphere. Hence we find
\begin{equation}
\label{5.21}\vc{\Omega}_\omega=\mathcal{Y}_R(\omega)\vc{N}_\omega,
\end{equation}
where $\mathcal{Y}_R(\omega)$ is the rotational admittance, given by
\begin{equation}
\label{5.22}\mathcal{Y}_R(\omega)=[-i\omega I_p+\zeta_R(\omega)]^{-1}.
\end{equation}

We consider in particular the applied torque
\begin{equation}
\label{5.23}\vc{N}(t)=\vc{L}\;\delta(t),
\end{equation}
where $\vc{L}$ is the imparted angular momentum. Correspondingly $\vc{N}_\omega=\vc{L}$. We define the rotational velocity relaxation function $\gamma_R(t)$ by
\begin{equation}
\label{5.24}\Omega(t)=\frac{L}{I^*}\;\gamma_R(t)\qquad t>0.
\end{equation}
with effective moment of inertia $I^*=I_p+I_a$, where $I_a$ is the added moment of inertia. The relaxation function is related to the admittance by
\begin{equation}
\label{5.25}\int^\infty_0e^{i\omega t}\gamma_R(t)\;dt=I^*\mathcal{Y}_R(\omega).
\end{equation}
Hence the relaxation function $\gamma_R(t)$ can be evaluated by inverse Fourier transform.
It has the properties
\begin{equation}
\label{5.26}\gamma_R(0+)=1,\qquad\int^\infty_0\gamma_R(t)\;dt=\frac{I^*}{\zeta_R(0)}.
\end{equation}
In general the added moment of inertia of the particle is defined by
\begin{equation}
\label{5.27}I_a=\lim_{\omega\rightarrow\infty}\frac{\zeta_R(\omega)}{-i\omega}.
\end{equation}
In the present case the added moment of inertia vanishes.

The net angular displacement of the sphere depends only on the initial angular momentum and the steady-state friction coefficient,
\begin{equation}
\label{5.28}\vc{\Delta}_R=\int^\infty_0\vc{\Omega}(t)\;dt=\vc{L}/\zeta_R(0).
\end{equation}
At low frequency the admittance has the expansion
\begin{equation}
\label{5.29}\mathcal{Y}_R(\omega)=\mu_R(0)+i\big(I_p\mu_R(0)^2+y_{R2}\big)\omega+\frac{1}{24\pi\eta}\;\alpha^3+O(\omega^2),
\end{equation}
where $\mu_R(0)=\zeta_{R0}^{-1}$ with $\zeta_{R0}=\zeta_R(0)$ is the steady-state mobility, and $y_{R2}=-\zeta_{R2}/\zeta_{R0}^2$. The third term gives rise to the long-time behavior
\begin{equation}
\label{5.30}\vc{\Omega}(t)\approx\frac{1}{\pi^{3/2}\rho(4\nu t)^{5/2}}\;\vc{L}\qquad\mathrm{as}\;t\rightarrow\infty.
\end{equation}
This depends only on the properties of the fluid.

It is convenient to define the dimensionless admittance $\hat{F}_R(x)$ with the complex variable $x=\alpha a$ as
\begin{equation}
\label{5.31}\hat{F}_R(x)=\frac{8\pi I_p}{15I_f}\;\eta a^3\mathcal{Y}_R(\omega),
\end{equation}
where $I_f=8\pi\rho a^5/15$ is the moment of inertia of displaced fluid, and we have chosen the pre-factor such that $\hat{F}_R(x)$ behaves as $1/x^2$ for large $x$. The function takes the form
\begin{equation}
\label{5.32}\hat{F}_R(x)=\frac{1}{x^2+M_RZ_R(x)},
\end{equation}
where we have abbreviated $M_R=15I_f/I_p$. For a rigid sphere with no-slip boundary condition the function takes the simple form\cite{1}
\begin{equation}
\label{5.32}\hat{F}_{Rns}(x)=\bigg[M_R+x^2+\frac{M_R x^2}{3+3x}\bigg]^{-1}.
\end{equation}

In Fig. 3 we show the real part of $\hat{F}_R(x)$ as a function of $\omega\tau_v$ for the same parameter values as before. The function shows a series of resonances. For comparison we also show the real part of $\hat{F}_{Rns}(x)$ with $M_R=15$. In Fig. 4 we show the corresponding relaxation function $\gamma_R(t)$ as a function of $t/\tau_v$, as obtained by numerical inverse Fourier transform. We compare with the function $\gamma_{Rns}(t)$ for the rigid no-slip sphere, given by a sum of three terms of the form Eq. (4.11). The function for the elastic sphere shows a spectacular sequence of vacillations.

\section{\label{V}Discussion}

We have shown that the velocity relaxation of an elastic sphere immersed in a viscous incompressible fluid has interesting behavior. In the short and intermediate time regime the velocity relaxation function differs strongly from that of a rigid sphere, both in translation and in rotation. The main results are contained in Eqs. (3.7) and (5.22). The relaxation functions are the Fourier transform of the complex admittance, which is determined by the frequency-dependent translational or rotational friction coefficient. The latter follow from the linearized equations of motion for fluid and solid, coupled via the no-slip boundary condition, and have been calculated explicitly.

The overlap of time scales of fluid and solid motion occurs for a sufficiently soft solid particle. This should be well within experimental feasibility. For gelatin Young's modulus is of the order $E\approx 20$ kPa \cite{13}. For water with viscosity $0.001$ Pa s this corresponds to a characteristic time $\tau_{VY}=\eta/E=5\times 10^{-8}$ s. For a particle of radius $a=1\;\mu\mathrm{m}$ the viscous relaxation time $\tau_v=\rho a^2/\eta=10^{-6}$ s is only slightly longer.

It will be of interest to study the quivering motion of a sufficiently elastic sphere at long times either in experiment or in computer simulation. The velocity autocorrelation function of Brownian motion can be studied experimentally in detail \cite{14}. It would be of interest to apply the same experimental techniques to the Brownian motion of an elastic particle.

For simplicity we have assumed a no-slip boundary condition, but the calculations can be performed also for a slip condition. Also the fluid can be made compressible \cite{7}. The account of particle elasticity can be extended to many particles. This is of theoretical interest for the modeling of the viscoelasticity of complex fluids \cite{15}.

The system studied here provides an interesting example of a vibrating mechanical continuum with an infinite set of discrete eigenfrequencies, coupled to open purely dissipative surroundings. As a consequence of the coupling the resonance spectrum becomes continuous with broadened peaks. After a simple excitation of the continuum it responds with a superposition of damped oscillations. The initial energy is dissipated in the surroundings. The initial momentum and angular momentum are imparted to the surroundings.

\newpage

\section*{Figure captions}

\subsection*{Fig. 1}
Plot of the real part of the translational reduced admittance $\hat{F}_T(x)$ as a function of frequency for a neutrally buoyant elastic sphere with choice of parameters given in Sec. 4 (solid curve). We compare with the same function for a no-slip rigid sphere for the same parameters, as given by Eq. (4.13) (dashed curve). At zero frequency $\hat{F}_T(0)=2/9$ for both curves.

\subsection*{Fig. 2}
Plot of the relaxation function $\gamma_T(t)$ as a function of $t/\tau_v$ for the same parameters as in Fig. 1 (solid curve) compared to the no-slip function $\gamma_{Tns}(t)$ (short dashes). We also compare with an approximate representation of the form Eq. (4.11) with a sum of eleven terms, as described in the text (long dashes).

\subsection*{Fig. 3}
Plot of the real part of the rotational reduced admittance $\hat{F}_R(x)$ as a function of frequency for a neutrally buoyant elastic sphere with choice of parameters given in Sec. 4 (solid curve). We compare with the same function for a no-slip rigid sphere for the same parameters, as given by Eq. (5.33) (dashed curve). At zero frequency $\hat{F}_R(0)=1/15$ for both curves.

\subsection*{Fig. 4}
Plot of the relaxation function $\gamma_R(t)$ as a function of $t/\tau_v$ for the same parameters as in Fig. 1 (solid curve) compared to the no-slip function $\gamma_{Rns}(t)$ (dashed curve).

\newpage
\setlength{\unitlength}{1cm}
\begin{figure}
 \includegraphics{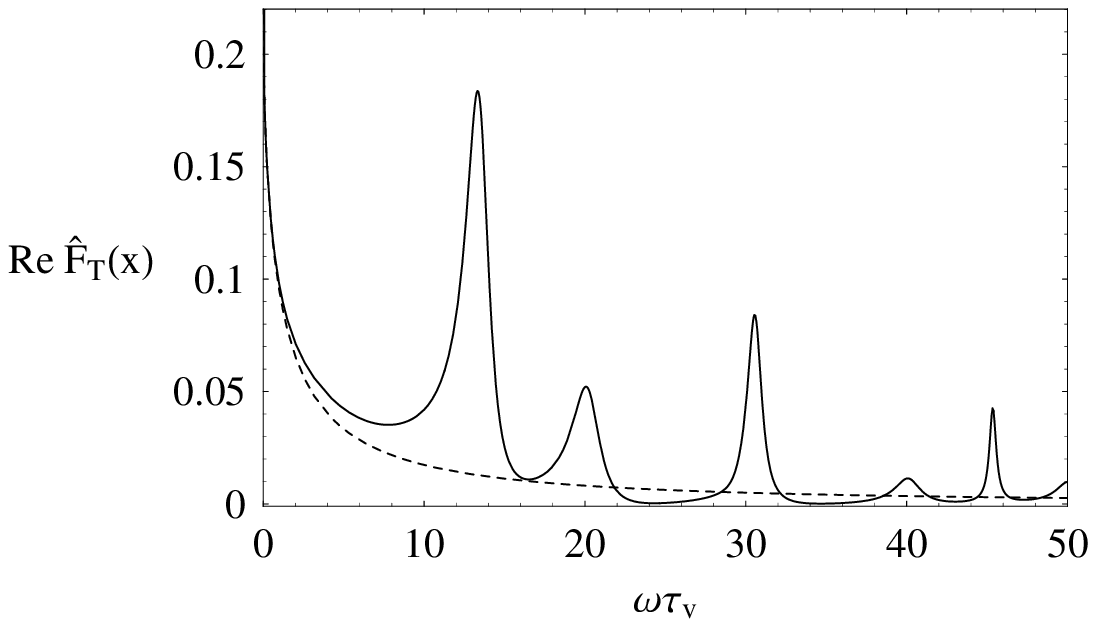}
   \put(-9.1,3.1){}
\put(-1.2,-.2){}
  \caption{}
\end{figure}
\newpage
\clearpage
\newpage
\setlength{\unitlength}{1cm}
\begin{figure}
 \includegraphics{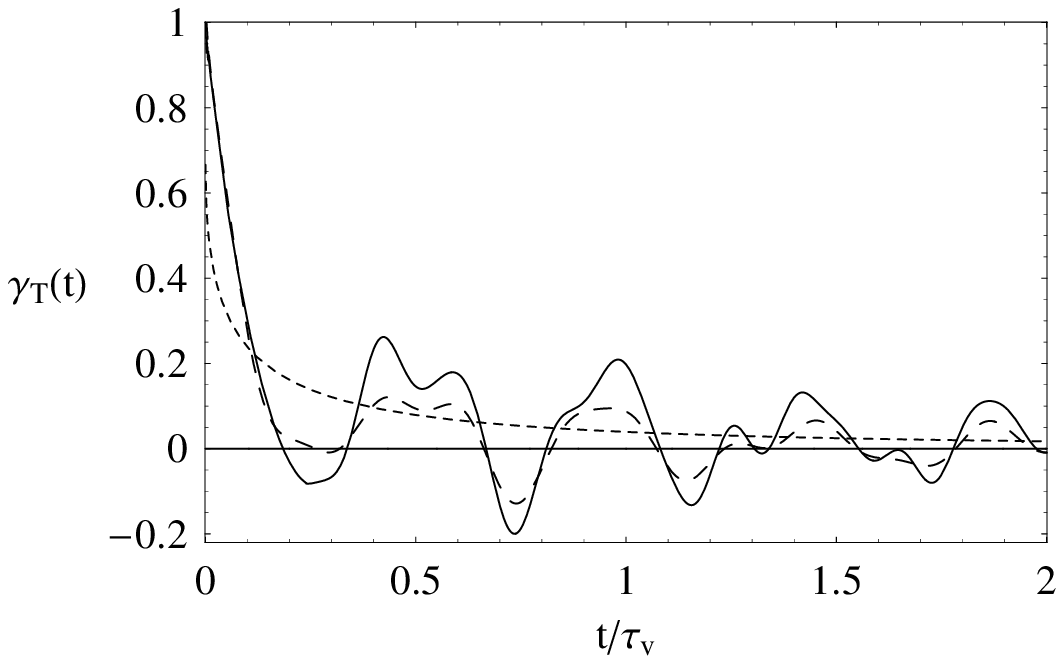}
   \put(-9.1,3.1){}
\put(-1.2,-.2){}
  \caption{}
\end{figure}
\newpage
\clearpage
\newpage
\setlength{\unitlength}{1cm}
\begin{figure}
 \includegraphics{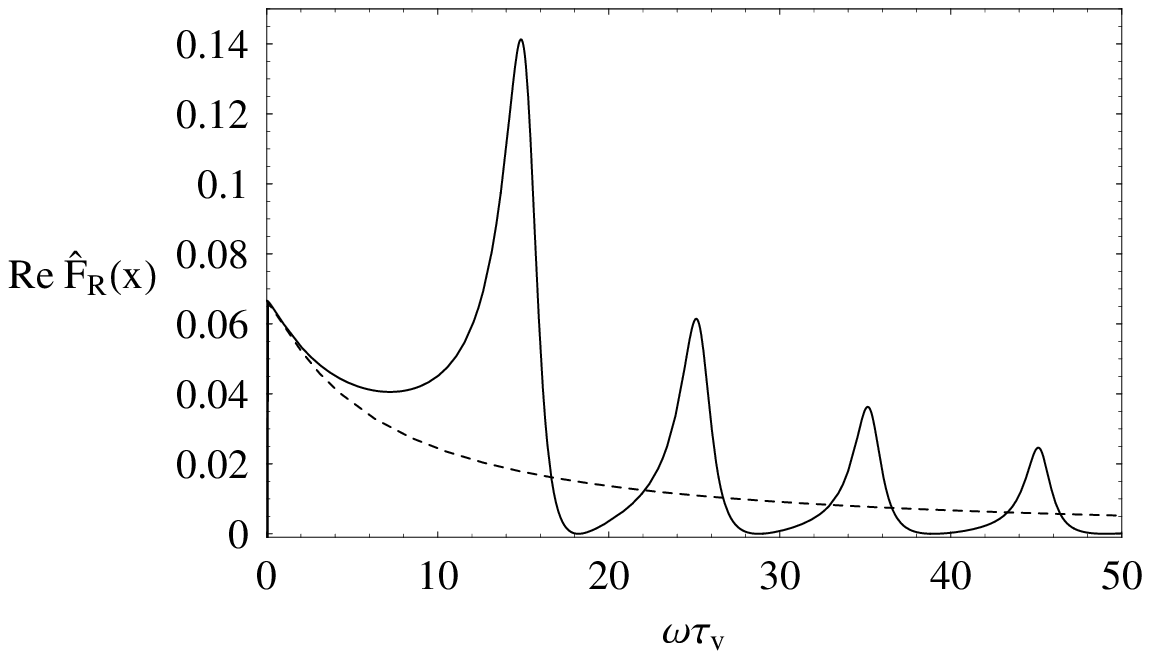}
   \put(-9.1,3.1){}
\put(-1.2,-.2){}
  \caption{}
\end{figure}
\newpage
\clearpage
\newpage
\setlength{\unitlength}{1cm}
\begin{figure}
 \includegraphics{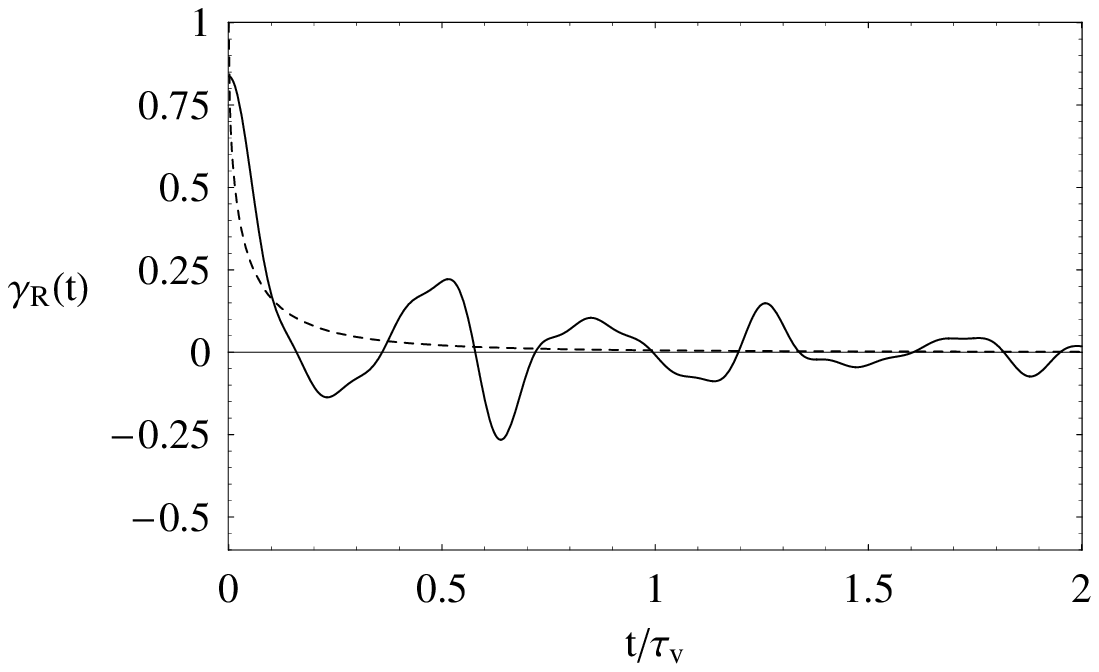}
   \put(-9.1,3.1){}
\put(-1.2,-.2){}
  \caption{}
\end{figure}
\newpage

\end{document}